\documentclass[10pt,conference]{IEEEtran}
\usepackage[T1]{fontenc}
\usepackage{mathtools}
\usepackage{xurl}
\usepackage{siunitx}
\usepackage{amsmath}
\usepackage{amssymb}
\usepackage{amsthm}
\usepackage{amsfonts}
\usepackage{algorithm} % for pseudo-code
\usepackage{algorithmic}
\usepackage{braket}

\usepackage{upgreek}
\usepackage{dirtytalk}

\usepackage{optidef}
\usepackage{cite}
\usepackage{authblk}
\usepackage{xcolor}

\newtheorem{lemma}{Lemma}
\newtheorem{definition}{Definition}
\usepackage{setspace}
\DeclarePairedDelimiter\abs{\lvert}{\rvert}%
\makeatletter
\let\oldabs\abs
\def\abs{\@ifstar{\oldabs}{\oldabs*}}

\begin{document}
\newcommand{\MC}[1]{\textcolor{blue}{Mahdi: #1}}
\newcommand{\BS}[1]{\textcolor{red}{Bernd: #1}}
\newcommand{\new} [1]{{\color{blue}#1}}
%General model
\newcommand{\timeIndex}{t}

%Users symbols
\newcommand{\numberOfUsers}{U}
\newcommand{\setOfUsers}{\mathcal{U}}
\newcommand{\userIndex}{i}
\newcommand{\requestIndex}{r}
\newcommand{\requestWithIndex}[1]{R_{#1}}
\newcommand{\minimumFidelity}{F_{\requestIndex}^{\text{min}}}

%QS symbols
\newcommand{\actionWithIndex}[1]{a_{#1}}
\newcommand{\actionIndex}{q}
\newcommand{\setOfActions}{\mathcal{A}}
\newcommand{\operationDistill}{a^{\text{dist}}}
\newcommand{\operationSwap}{a^{\text{swap}}}

%Matching symbols
\newcommand{\userPreferenceWithIndex}[1]{\succ^{\mathrm{Req}}_{#1}}
\newcommand{\userPreference}{\userPreferenceWithIndex{\requestIndex}}
\newcommand{\elementOfAssignmentMatrix}{\elementOfAssignmentMatrixWithIndex{\requestIndex,\actionIndex,\timeIndex}}

\newcommand{\QSPreferenceWithIndex}[1]{\succ^{\mathrm{QS}}_{#1}}
\newcommand{\QSPreference}{\QSPreferenceWithIndex{\actionIndex}}
\newcommand{\setOfAssociatedRequests}{\mathcal{R}_q^{\eta}}
\newcommand{\setOfAssociatedRequestsPrime}{\mathcal{R}_q^{\eta'}}
\newcommand{\setOfAssociatedRequestsQPrimeEtaPrime}{\mathcal{R}_{q'}^{\eta'}}
\newcommand{\setOfAssociatedRequestsQPrime}{\mathcal{R}_{q'}^{\eta}}
\newcommand{\setOfAssociatedRequestsWithIndices}[2]{\mathcal{R}_{#1}^{#2}}

\newcommand{\elementOfAssignmentMatrixWithIndex}[1]{x_{#1}}
\newcommand{\assignmentMatrix}{\boldsymbol{X}_{\timeIndex}}

\newcommand{\request}{r_l^{k,m}}

\title{Matching Game for Optimized Association in Quantum Communication Networks}

\author[1]{Mahdi Chehimi}
\author[2]{Bernd Simon}
\author[1]{Walid Saad}
\author[2]{Anja Klein}
\author[3]{Don Towsley}
\author[4]{M\'erouane Debbah}
\affil[1]{\small Wireless@VT, Bradley Department of Electrical and Computer Engineering, Virginia Tech, Arlington, VA USA}
\affil[2]{\small Communication Engineering Lab, Technische Universit\"{a}t Darmstadt, Darmstadt, Germany}
\affil[3]{\small University of Massachusetts Amherst, Amherst, MA USA}
\affil[4]{\small Technology Innovation Institute, 9639 Masdar City, Abu Dhabi, United Arab Emirates}
\affil[ ]{\small \textit{\{mahdic,walids\}@vt.edu}, \textit{\{b.simon, a.klein\}@nt.tu-darmstadt.de}, \textit{towsley@cs.umass.edu}, \textit{merouane.debbah@tii.ae}\vspace{-0.085in}}

% \thanks{M. Debbah is with the Technology Innovation Institute, 9639 Masdar City, Abu Dhabi, United Arab Emirates,
% and also with CentraleSupelec, University Paris-Saclay, 91192 Gif-sur-Yvette, France
% Email: merouane.debbah@tii.ae.}

\maketitle

\begin{abstract}
Enabling quantum switches (QSs) to serve requests submitted by quantum end nodes in quantum communication networks (QCNs) is a challenging problem due to the heterogeneous fidelity requirements of the submitted requests and the limited resources of the QCN. Effectively determining which requests are served by a given QS is fundamental to foster developments in practical QCN applications, like quantum data centers. However, the state-of-the-art on QS operation has overlooked this association problem, and it mainly focused on QCNs with a single QS. In this paper, the request-QS association problem in QCNs is formulated as a matching game that captures the limited QCN resources, heterogeneous application-specific fidelity requirements, and scheduling of the different QS operations. To solve this game, a \emph{swap-stable} request-QS association (RQSA) algorithm is proposed while considering partial QCN information availability. Extensive simulations are conducted to validate the effectiveness of the proposed RQSA algorithm. Simulation results show that the proposed RQSA algorithm achieves a near-optimal (within $5\%$) performance in terms of the percentage of served requests and overall achieved fidelity, while outperforming benchmark greedy solutions by over $13\%$. Moreover, the proposed RQSA algorithm is shown to be scalable and maintain its near-optimal performance even when the size of the QCN increases.
\end{abstract}

\IEEEpeerreviewmaketitle
\vspace{-0.1in}
\section{Introduction}
\vspace{-0.05in}
Quantum communication networks (QCNs) are seen as a pillar of future communication technologies due to their advantages in terms of security, sensing capabilities, and computational powers. QCNs rely on the creation and distribution of Einstein-Podolsky-Rosen (EPR) pairs of \emph{entangled quantum states} between distant QCN nodes \cite{chehimi2022physics}. Each EPR pair consists of two inherently-correlated photons, each of which is transferred to a QCN node to establish an end-to-end (e2e) entangled connection. However, the fragile nature of entangled photons results in exponential losses that increase with the travelled distance over quantum channels, e.g., optical fiber. As such, intermediate quantum repeater nodes are needed to split long distances into shorter segments by performing \emph{entanglement swapping} on entangled photons to connect distant QCN nodes \cite{briegel1998quantum}. When such repeaters share multiple EPR pairs with several QCN nodes to create e2e connections, they are called \emph{quantum switches (QSs)}.

In practice, a QS has a limited-capacity quantum memory for photon storage. A heralding station is responsible for generating EPR pairs and distributing each pair between the QS and other QCN nodes to create link-level connections (LLCs). The \emph{fidelity}, or quality, of an LLC can be enhanced by performing \emph{entanglement distillation} before swapping two LLCs to establish an e2e connection \cite{bennett1996purification}. Practical applications, like quantum data centers and quantum cloud networks, encompass QCN setups with multiple QSs connecting several end-node quantum devices. The design of such multiple-QS QCNs requires overcoming many challenges such as the limited storage capacity of QSs, imperfections associated with EPR generation and transmission, the need to schedule the different QS operations (i.e., entanglement swapping and distillation), and the presence of heterogeneous application-specific minimum fidelity requirements.

Multiple prior works \cite{zhao2021redundant,pant2019routing,vardoyan2019stochastic,vasantam2021stability,dai2021entanglement,panigrahy2022capacity,promponas2023full} attempted to address some of the aforementioned challenges, and those works can be divided into three main types. First, some works, like \cite{zhao2021redundant}, considered a QS-based multi-hop QCN and performed entanglement provisioning and \emph{path selection} to maximize throughput. Second, prior works, such as \cite{pant2019routing}, considered \emph{routing} EPR pairs over several QCN paths to create e2e connections. The last type, which is the most relevant to our work, considered star-shaped QCNs, where several nodes are connected to a single QS through EPR pairs \cite{vardoyan2019stochastic,vasantam2021stability,dai2021entanglement,panigrahy2022capacity,promponas2023full}. For instance, the work in \cite{vardoyan2019stochastic} was the first to consider aggregate QS capacity and analytically analyze its stability. However, almost-perfect conditions were assumed in \cite{vardoyan2019stochastic}. Additionally, the work in \cite{vasantam2021stability} considered a QCN with a QS serving requests having minimum fidelity constraints. However, entanglement distillation was not considered in \cite{vasantam2021stability}. Meanwhile, the work in \cite{dai2021entanglement} studied QS stability and swap scheduling. However, the authors in \cite{dai2021entanglement} did not include entanglement distillation and assumed an infinite lifetime of EPR pairs. Moreover, the work in \cite{panigrahy2022capacity} analyzed the capacity regions and stability of a single QS and scheduled swapping/distillation operations to satisfy minimum fidelity requirements while considering noisy gates and measurements. However, \cite{panigrahy2022capacity} considered a homogeneous fidelity for all link-level EPR pairs. Finally, the authors in \cite{promponas2023full} proposed a memory allocation policy for a constrained QS operation in a star-shaped QCN. However, the model proposed in \cite{promponas2023full} did not account for fidelity requirements of both link-level and e2e connections and did not schedule distillation operations. %Additionally, it only considered a maximum of one request of each QCN e2e type,

%Similar simplified assumptions were made in \cite{dai2020quantum}, where the authors proposed a control policy for scheduling the quantum memory of a central QS.

%Such prior works investigate some operational questions about a QS, where end nodes send requests for establishing e2e connections to the QS, which performs scheduling of its operations and allocation of its memory slots to serve the submitted requests. 

% considering the QSs' resources, fidelity, and scheduling of their operations in a joint manner. 

% considered a single QS and did not perform a joint analysis of QSs where entanglement distillation capabilities, are considered, along with memory limitations and swap/distillation scheduling. 

Furthermore, these prior works \cite{vardoyan2019stochastic,vasantam2021stability,dai2021entanglement,panigrahy2022capacity,promponas2023full} focused on a single QS handling all e2e requests and did not consider multiple QSs connected to several end nodes with heterogeneous resources and fidelity constraints. In such a QCN setup (see Fig.~\ref{fig:system_model}), it is essential to \emph{associate} each request with the QS that optimizes its fidelity. This \emph{request-QS association problem}, which is essential for designing quantum data centers, has been overlooked in prior works. Accordingly, there is a need for a thorough investigation of the request-QS association problem in QCNs with multiple QSs while taking into consideration the scheduling of QS entanglement swapping/distillation operations, memory limitations, and performance requirements. 

The main contribution of this work is a novel matching-based framework for optimizing request-QS association in QCNs with multiple QSs, possessing heterogeneous resources, that satisfy QCN users' performance requirements while considering practical constraints of QCN elements. To the best of our knowledge, this is the first work to explore this research area, and therefore, we make the following key contributions:
\begin{itemize}
    \item We propose the first holistic analysis of the request-QS association problem in QCNs under limited resource constraints and heterogeneous fidelity requirements. 
    
    \item We formulate the request-QS association problem as a \emph{matching game}~\cite{gale1962matching} where both requests and QSs rank each other based on fidelity-maximizing preferences. This novel matching approach enables us to solve the considered association problem without requiring full knowledge of QCN information.

    \item We propose a novel \emph{request-QS association (RQSA)} algorithm based on \emph{swap-matching}~\cite{BodineBaron2011swapMatching} to solve the formulated matching game while guaranteeing convergence under partial QCN information availability. 
    
    \item Simulation results show that our RQSA algorithm is scalable and achieves a near-optimal performance within $5\%$ of the optimal solution in terms of served requests and overall served e2e fidelity. 
\end{itemize}
\vspace{-0.08in}
\section{System Model}\vspace{-0.02in}
\label{sec_system_model}

\begin{figure}[t!]
\begin{center}%\vspace{-0.25in}
\centerline{\includegraphics{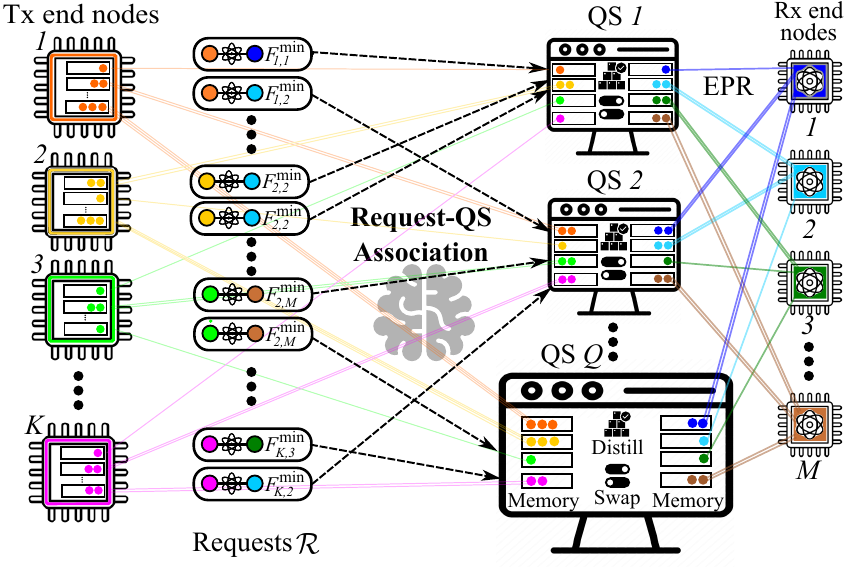}}\vspace{-0.17in}
\caption{Studied QCN model for the requests-QSs association problem.}
\label{fig:system_model}
\end{center}\vspace{-0.33in}
\end{figure}

Consider a QCN composed of a set $\mathcal{Q}$ of $Q$ QSs connected to a set of end nodes through link-level EPR pairs. The end nodes are split into transmitting (Tx) and receiving (Rx) nodes, where requests for e2e connections are sent from Tx nodes to the QSs (see Fig.~\ref{fig:system_model}). Moreover, $\mathcal{K}$ denotes the set of $K$ Tx nodes, and $\mathcal{M}$ the set of $M$ Rx nodes. 

The operation of the QCN occurs in a time-slotted manner. Prior to each time slot, heralding stations installed between QSs and end nodes attempt to create $n$ link-level EPR pairs to connect every QS $q\in\mathcal{Q}$ to every Tx (and Rx) node, $k\in\mathcal{K}$ (and $m\in\mathcal{M}$), respectively, with a probability of success $p_{k,q}$ (and $p_{q,m}$) for each pair that depends on the corresponding link length. Accordingly, the link-level EPR generation process between a QS and a Tx (or Rx) node follows a \emph{binomial distribution} with parameters $n$ and $p_{k,q}$ (or $p_{q,m}$) \cite{panigrahy2022capacity}. Thus, each QS is connected to all Tx nodes through $N^{\text{Tx}}_{k,q}$ successfully-generated link-level EPR pairs each having a \emph{fidelity} of $F^{\text{Tx}}_{k,q}$. Similarly, every QS is linked to each Rx node through $N^{\text{Rx}}_{q,m}$ successfully-generated link-level EPR pairs each of fidelity $F^{\text{Rx}}_{q,m}$. The EPR pairs are then stored in quantum memories at both the QSs and end nodes. Those pairs are assumed to remain coherent and maintain their fidelities for one time slot, before being discarded. 

At the beginning of each time slot, Tx nodes submit a set $\mathcal{R}$ of $R$ requests to the QSs. Each request is represented as a tuple, $r_l^{i,j} = (i,j,F_{i,j}^{\text{min}})$, where $l\in \{1,2,...,R\}$, $i\in\mathcal{K}$, and $j\in\mathcal{M}$. Here, $r_l^{k,m}$ represents a request by Tx node $k\in\mathcal{K}$ to establish a single e2e EPR pair with Rx node $m\in\mathcal{M}$ with a minimum fidelity of $F_{k,m}^{\text{min}}$. In addition, each submitted request must be served, if feasible, during its submission time slot, or be discarded. We assume that, during each time slot, every Tx node may submit multiple repeated requests that are identical and have exactly the same required minimum fidelity, since they intend to serve the same application.

In our model, we consider that only partial QCN information is available to the Tx nodes when submitting their requests. In particular, each Tx node has access to only the information related to its link-level EPR pairs with every QS. Moreover, the QSs publicly announce information about their link-level EPR pairs with every Rx node to the Tx nodes.

Each QS $q\in\mathcal{Q}$ can perform two distinct quantum operations: 1) \emph{entanglement swapping}, to connect a Tx node to an Rx node, and 2) \emph{entanglement distillation} to enhance the fidelity of link-level EPR pairs. Every link-level EPR pair is represented by a Werner state $\rho = W\ket{\psi_{00}}\bra{\psi_{00}} + \frac{1-W}{4}\Pi,$ where $W$ is the Werner parameter that directly affects the fidelity of those pairs, which is given as: $F = \frac{3W+1}{4}$ \cite{cope2018converse}.

When a QS $q\in\mathcal{Q}$ swaps two link-level EPR pairs, one with Tx node $k\in\mathcal{K}$ having fidelity $F^{\text{Tx}}_{k,q}$, and the other with Rx node $m\in\mathcal{M}$ having fidelity $F^{\text{Rx}}_{q,m}$, the resulting e2e EPR pair has a fidelity given by \cite{briegel1998quantum}: \vspace{-0.05in}\begin{equation}\label{eq_nested_swaps_unequal}\footnotesize
    S(F^{\text{Tx}}_{k,q},F^{\text{Rx}}_{q,m}) = \frac{1}{4} + \frac{3}{4}\biggl(\frac{4F^{\text{Tx}}_{k,q}-1}{3}\biggl)\biggl(\frac{4F^{\text{Rx}}_{q,m}-1}{3}\biggl).
\end{equation}

We adopt the Oxford entanglement distillation protocol \cite{bennett1996purification} for performing entanglement distillation of two link-level EPR pairs. According to this protocol, two identical EPR pairs with initial fidelity $F_{\text{initial}}$ can be distilled into one EPR pair having a higher fidelity given by \cite{bennett1996purification}:

\noindent\begin{equation}\label{eq_fidelity_distillation_identical}\footnotesize
    D(F_{\text{initial}}) = \frac{(F_{\text{initial}})^2 + (\frac{1-F_{\text{initial}}}{3})^2}{(F_{\text{initial}})^2 + 2F_{\text{initial}}(\frac{1-F_{\text{initial}}}{3}) + 5(\frac{1-F_{\text{initial}}}{3})^2}.
\end{equation}

To simplify the analysis, a QS is assumed to perform at most one distillation operation for each link-level EPR pair. Also, if performed, distillation is considered to always precede entanglement swapping \cite{panigrahy2022capacity}. Accordingly, there are four possible actions regarding the scheduling of the entanglement swapping/distillation operations to handle each submitted request that every QS can take.\footnote{A higher number of possible actions can be easily integrated into our model by allowing QSs to perform more distillation operations. Also, due to space constraints, we have omitted the terms corresponding to measurement errors and gate noise from the given swapping and distillation fidelity expressions, which also can be easily integrated into our model.} The action choice directly affects the fidelities of the resulting e2e EPR pairs and the number of available link-level EPR pairs in quantum memories. Here, we introduce $\alpha_j^{\mathrm{Tx}}$ and $\alpha_j^{\mathrm{Rx}}$ to denote the number of \emph{utilized link-level EPR pairs} from both Tx and Rx nodes' memories, respectively, as a result of each possible QS action $j\in\{1,2,3,4\}$. The four considered actions and their corresponding impacts are: 

\subsubsection{Direct entanglement swapping} Swap one link-level EPR pair connected to Tx node $k\in\mathcal{K}$ with one link-level EPR pair connected to Rx node $m\in\mathcal{M}$. When QS $q\in\mathcal{Q}$ performs this action to serve request $r_l^{k,m}$, the fidelity of the resulting e2e EPR pair will be $F_{q,k,m,1}^{\text{e2e}} = S(F^{\text{Tx}}_{k,q},F^{\text{Rx}}_{q,m})$. Consequently, the number of link-level EPR pairs between QS $q\in\mathcal{Q}$ and Tx node $k\in\mathcal{K}$ and Rx node $m\in\mathcal{M}$, i.e., $N_{k,q}^{\text{Tx}}$ and $N^{\text{Rx}}_{q,m}$, respectively, are both reduced by 1. The number of utilized link-level EPR pairs associated with the \emph{direct entanglement swapping} action are given by $\alpha_1^{\mathrm{Tx}} = \alpha_1^{\mathrm{Rx}} = 1$.

\subsubsection{Tx distillation, then entanglement swapping} Distill two link-level EPR pairs connected to the Tx node $k\in\mathcal{K}$, then swap the distilled pair with an EPR pair connected to the Rx node $m\in\mathcal{M}$. When QS $q\in\mathcal{Q}$ performs this action to serve a request $r_l^{k,m}$, the fidelity of the resulting e2e EPR pair is $F_{q,k,m,2}^{\text{e2e}} = S(D(F^{\text{Tx}}_{k,q}),F^{\text{Rx}}_{q,m})$. Consequently, the number of link-level EPR pairs between QS $q\in\mathcal{Q}$ and Tx node $k\in\mathcal{K}$ is reduced by 2, while the number of link-level EPR pairs between QS $q\in\mathcal{Q}$ and Rx node $m\in\mathcal{M}$ is reduced by 1, as the entanglement distillation utilizes two link-level EPR pairs. The number of utilized link-level EPR pairs associated with the \emph{Tx distillation, then entanglement swapping} action are $\alpha_2^{\mathrm{Tx}} = 2$, and $\alpha_2^{\mathrm{Rx}} = 1$.
 
\subsubsection{Rx distillation, then entanglement swapping} Distill two link-level EPR pairs connected to Rx node $m\in\mathcal{M}$, then swap the distilled pair with an EPR pair connected to Tx node $k\in\mathcal{K}$. When QS $q\in\mathcal{Q}$ performs this action to serve request $r_l^{k,m}$, the fidelity of the resulting e2e EPR pair is $F_{q,k,m,3}^{\text{e2e}} = S(F^{\text{Tx}}_{k,q},D(F^{\text{Rx}}_{q,m}))$. Consequently, the number of link-level EPR pairs between QS $q$ and Tx node $k$ is reduced by 1, while the number of link-level EPR pairs between QS $q\in\mathcal{Q}$ and Rx node $m$ is reduced by 2. The number of utilized link-level EPR pairs associated with the \emph{Rx distillation, then entanglement swapping} action are $\alpha_3^{\mathrm{Tx}} = 1$, and $\alpha_3^{\mathrm{Rx}} = 2$.
 
\subsubsection{Tx \& Rx distillation, then entanglement swapping} Distill two link-level EPR pairs connected to Tx node $k\in\mathcal{K}$, and simultaneously distill two EPR pairs connected to Rx node $m\in\mathcal{M}$, then swap the two distilled pairs. When QS $q\in\mathcal{Q}$ performs this action to serve request $r_l^{k,m}$, the fidelity of the resulting e2e EPR pair is $F_{q,k,m,4}^{\text{e2e}} = S(D(F^{\text{Tx}}_{k,q}),D(F^{\text{Rx}}_{q,m}))$. Consequently, the number of link-level EPR pairs between QS $q$ and Tx node $k$ and Rx node $m$ are both reduced by 2. The numbers of utilized link-level EPR pairs associated with the \emph{Tx \& Rx distillation, then entanglement swapping} action are $\alpha_4^{\mathrm{Tx}} = \alpha_4^{\mathrm{Rx}} = 2$.

To simplify notation, we introduce the vectors $\boldsymbol{\alpha}^{\mathrm{Tx}} = [1,2,1,2]^{\mathrm{T}}$ and $\boldsymbol{\alpha}^{\mathrm{Rx}} = [1,1,2,2]^{\mathrm{T}}$ of \emph{utilized link-level EPR pairs} that result from the four possible QS actions. Next, we formulate the request-QS association problem and propose a matching game formulation~\cite{roth1990matchingTheory}.
\vspace{-0.05in}
\section{Request-QS Association as a Matching Game}
\subsection{Request-QS Association Problem}
In the request-QS association problem, a submitted request $r_l^{k,m}\in\mathcal{R}$ must be associated, if feasible, with at most one QS $q\in\mathcal{Q}$, or be discarded. This QS performs one of the four aforementioned actions to serve the request during a time slot. 
%Let $\Lambda = \mathcal{R} \times \mathcal{Q}$ denote the set of all possible associations between requests and QSs.
We define \emph{matching} $\eta$ as an association between QSs and requests. 
The association between a submitted request $\request$ and a QS $q$ is denoted as $(\request,q) \in \eta$.
Each QS $q\in\mathcal{Q}$ can serve multiple requests. 
We define $\setOfAssociatedRequests \subseteq \mathcal{R}$ as the set of requests associated with QS $q$ in matching $\eta$.
As multiple requests are associated with each QS, we have a \emph{many-to-one} matching problem.

Each submitted request must be served with the highest fidelity possible.
Therefore, we define the utility of a submitted request $r_l^{k,m}\in\mathcal{R}$ when associated with QS $q\in\mathcal{Q}$ as the \emph{fidelity} of its generated e2e EPR pair:\vspace{-0.05in}
\begin{align}\label{eq_request_tility}
    U_{l}(q) = F_{q,k,m,1}^{\text{e2e}} = S(F^{\text{Tx}}_{k,q},F^{\text{Rx}}_{q,m}),
\end{align}where $k\in\mathcal{K}$ and $m\in\mathcal{M}$ are the corresponding Tx/Rx nodes, respectively, in $r_l^{k,m} = (k,m,F_{k,m}^{\text{min}})$. (\ref{eq_request_tility}) considers the worst case for QS $q$, which corresponds to taking the \emph{direct entanglement swap} action without any distillation, since that action yields the lowest fidelity of the resulting e2e EPR pair. This worst-case assumption stems from the fact that the request (i.e., the end node) does not know which action will be taken by its prospective QS $q$. 

Similarly, each QS aims to serve each request with the highest fidelity possible. In matching $\eta$, for each individual request $r_l^{k,m}\in\mathcal{R}$ served by QS $q\in\mathcal{Q}$, the respective QS utility for that request is the resulting e2e EPR pair's fidelity:
\begin{equation}\label{eq_QS_single_request_utility}\small
\tilde U_q(r_l^{k,m}) =
\begin{cases}
F^{\text{e2e}}_{q,k,m,j_{q,k,m}(\eta)}, \text{  if  } F^{\text{e2e}}_{q,k,m,j_{q,k,m}(\eta)}\geq F^{\text{min}}_{k,m}\\
-\infty, \quad\quad\quad\quad \text{else}.
\end{cases}
\end{equation}

In \eqref{eq_QS_single_request_utility}, $j_{q,k,m}(\eta)$ captures the fact that the fidelity of the resulting e2e EPR pair depends on the action taken by the QS. For instance, $j_{q,k,m}(\eta) \in \{1,2,3,4\}$ represents the action taken by the QS to serve request $r_l^{k,m}$ based on matching $\eta$. The second case in the above expression corresponds to the situation when the QS cannot serve the request because it cannot provide the request's minimum fidelity requirement. % of the submitted request.

Each QS $q\in\mathcal{Q}$ must decide on the actions that maximize the fidelity for its associated requests, i.e., maximize (\ref{eq_QS_single_request_utility}) for each request. After the optimal actions are identified, the overall utility of QS $q\in\mathcal{Q}$ for its associated set of requests $\setOfAssociatedRequests$ in matching $\eta$ is the sum of the individual request utilities:
\vspace{-0.1cm}
\begin{equation}
\label{eq:QSutility}
    U_q(\setOfAssociatedRequests) = \sum_{r_l^{k,m}\in\setOfAssociatedRequests} \tilde U_q(r_l^{k,m}),
    \vspace{-0.1cm}
\end{equation}
which captures the fact that the goal of each QS $q \in \mathcal{Q}$ is to maximize the overall delivered e2e fidelities for the set of associated requests $\setOfAssociatedRequests$. 

The process of selecting the actions to serve the associated requests in $\setOfAssociatedRequests$ by QS $q\in\mathcal{Q}$ can be formulated as an optimization problem. To do so, we define $\boldsymbol{A}_q$ as the actions matrix for QS $q$, which includes all possible actions for all its associated requests $\request\in\setOfAssociatedRequests$. In particular, $\boldsymbol{A}_q = [\boldsymbol{a}_1,\boldsymbol{a}_2,\boldsymbol{a}_3,\boldsymbol{a}_4]$, where each vector $\boldsymbol{a}_j$ is of dimension $\abs{\setOfAssociatedRequests}\times 1$, and each entry $a_{l,j}$ of $\boldsymbol{a}_j$, given $l\in\{1,2,...,\abs{\setOfAssociatedRequests}\}$ and $j\in\{1,2,3,4\}$, corresponds to a request $r_l^{k,m}\in\setOfAssociatedRequests$. Each element $a_{l,j}$ is binary, where it takes a value of one when action $j$ is performed to serve request $r_l^{k,m}$. The dimension of $\boldsymbol{A}_q$ is $\abs{\setOfAssociatedRequests}\times 4$. Accordingly, the action-selection optimization problem for QS $q\in\mathcal{Q}$ is:

\begin{subequations}
\vspace{-0.35cm}
\small
\begin{alignat}{2}
\mathcal{P}1: \quad &\!\underset{\boldsymbol{A}_q}{\max}        &\quad& \sum_{r_l^{k,m}\in\setOfAssociatedRequests} U_q(r_l^{k,m}) \label{eq:optProb}\\
&   s.t.               &      & \sum_{i:r_i = r_i^{k,m}, \forall m\in\mathcal{M}_q} \mkern-18mu \boldsymbol{A}_q\cdot\boldsymbol{\alpha}^{\mathrm{Tx}} \leq N^{\text{Tx}}_{k,q}, \quad \forall k\in\mathcal{K}_q\label{eq:constraint1},\\
&                  &      & \sum_{i:r_i = r_i^{k,m}, \forall k\in\mathcal{K}_q} \mkern-18mu  \boldsymbol{A}_q\cdot\boldsymbol{\alpha}^{\mathrm{Rx}} \leq N^{\text{Rx}}_{q,m}, \quad \forall m\in\mathcal{M}_q\label{eq:constraint2},
\end{alignat}
\end{subequations}
where the objective function corresponds to the overall utility achieved by QS $q\in\mathcal{Q}$ from all its associated requests $\setOfAssociatedRequests$. Constraint \eqref{eq:constraint1} ensures that the number of used link-level EPR pairs between the QS and Tx node $k\in\mathcal{K}_q$ does not exceed the number of available link-level EPR pairs between them, $N^{\text{Tx}}_{k,q}$, $\forall k,q\in\mathcal{K}_q,\mathcal{Q}$. Similarly, constraint \eqref{eq:constraint2} ensures that the number of consumed link-level EPR pairs between the the QS and Rx node $m\in\mathcal{M}_q$ does not exceed the number of available link-level EPR pairs between them, $N^{\text{Rx}}_{q,m}$, $\forall q,m\in\mathcal{Q},\mathcal{M}_q$.

Solving the request-QS association problem is challenging, because it must factor in the limited number of available link-level EPR pairs of the Tx and Rx nodes and the QSs. 
Also, each QS must schedule its actions such that the maximum number of submitted requests in the QCN is served during each time step.
Solving the request-QS association problem using classical optimization techniques is impractical because the number of possible combinations of associated requests per QS is $2^R$, i.e., the complexity grows exponentially with $R$. 
Accordingly, we propose a computationally efficient, decentralized approach that accounts for the partial QCN information availability. 

\vspace{-0.1in}
\subsection{Matching Game Formulation}
Matching theory \cite{roth1990matchingTheory} is a powerful tool that has been adopted to solve several complex communication network problems \cite{chen2021matching}. Here, we leverage matching theory to formulate the request-QS association problem as a matching game so as to overcome its exponentially growing complexity. 
Note that our formulation differs from prior works on matching games for classical wireless systems~\cite{gu2015matchingTheory} in the fact that we have to consider quantum-specific constraints regarding the fidelity of EPR pairs, limited quantum memory, and heterogeneous minimum fidelity requirements. Formally, the proposed matching game is defined as follows.
\begin{definition}[Matching game]
\label{def:matching_game}
A matching game is defined by two sets of matching parties ($\mathcal{R}, \mathcal{Q}$) and two preference relations $\userPreference$, $\QSPreference$ allowing each submitted request $\request \in \mathcal{R}$ to rank the QSs and each QS $q \in \mathcal{Q}$ to rank sets of associated requests.
\end{definition} 
For any request $\request$, a \emph{preference relation} $\userPreference$ is defined over the set of QSs $\mathcal{Q}$ such that, for any two QSs, $q,q'\in\mathcal{Q}$, we have:\vspace{-0.1in}\begin{align}
q \userPreference q' \Leftrightarrow U_{l}(q) > U_{l}(q'),
\label{eq:requestPreferences}
\end{align}\vspace{-0.18in} 

\noindent which means that request $r_l^{k,m}$ prefers QS $q\in\mathcal{Q}$ over QS $q'\in\mathcal{Q}$ whenever the utility~\eqref{eq_request_tility} associated with $q\in\mathcal{Q}$ is higher than the utility associated with $q'$.

Similar to the case of requests, for any QS $q\in\mathcal{Q}$, we define a \emph{preference relation} $\QSPreference$ over the set of associated requests $\setOfAssociatedRequests$.
For any two matchings $\eta,\eta'$, the QS ranks the corresponding sets of associated requests $\setOfAssociatedRequests$ in matching $\eta$ and $\setOfAssociatedRequestsPrime$ in matching $\eta'$ as follows:\vspace{-0.05in}
\begin{align}
\label{eq:QSpreferences}
\setOfAssociatedRequests \QSPreference \setOfAssociatedRequestsPrime \Leftrightarrow U_{q}(\setOfAssociatedRequests) > U_{q}(\setOfAssociatedRequestsPrime),
\end{align}
which means that the QS $q\in\mathcal{Q}$ prefers the set $\setOfAssociatedRequests$ of associated requests in matching $\eta$ over the set $\setOfAssociatedRequestsPrime$ in matching $\eta'$ whenever the overall utility~\eqref{eq:QSutility} associated with $\setOfAssociatedRequests$ is higher than the overall utility associated with $\setOfAssociatedRequestsPrime$.

\vspace{-0.07in}
\subsection{Proposed Solution and Algorithm}\vspace{-0.02in}
In this section we propose an algorithm to find a \textit{stable} matching $\eta$.
Classical definitions of stability~\cite{gale1962matching,roth1990matchingTheory} in matching games, which rely on preferences of individual matching parties, cannot be applied to our proposed matching formulation. This is because the preference relations~\eqref{eq:QSpreferences} require QSs to rank sets of associated requests instead of individual requests.
To overcome this challenge, we adopt the definition of \textit{swap stability}~\cite{BodineBaron2011swapMatching}, which means that no submitted request or QS can increase its utility by swapping its current matching partner.\footnote{Note that swap stability is not to be confused with the entanglement swap operation that a QS can perform on two link-level EPR pairs.}
The foundation for the analysis of swap stability is a \emph{swap matching}, which simply results from two requests $r$ and $r'$ exchanging their respective associated QSs $q$ and $q'$ in $\eta$.
Formally, given a matching $\eta$, two submitted requests $r,r'\in\mathcal{R}$ and two QSs $q,q'\in\mathcal{Q}$ with $(r,q), (r',q') \in \eta$, a swap matching is defined as $\eta_{r,r'}^{q} = \eta \setminus \{(r,q) , (r',q')\} \cup \{(r',q), (r,q')\}$. 
Accordingly, swap stability is defined as:
\begin{definition}[Swap stability] 
\label{def:swap_stability}
A matching $\eta$ is said to be swap stable if no swap matching $\eta' = \eta_{r,r'}^{q}$ exists such that:\\
(i) Request $r$ and $r'$ prefer the QSs associated in swap matching $\eta'$ over the QSs associated in $\eta$. Formally, both prefer to swap their respective QSs $q' \userPreference q$ and $q \userPreferenceWithIndex{r'} q'$, and\\
(ii) QS $q\in\mathcal{Q}$ and $q'$ prefer the requests associated in swap matching $\eta'$ over the requests associated in $\eta$. Formally, $\setOfAssociatedRequestsWithIndices{q}{\eta'} \QSPreference \setOfAssociatedRequests$ and $\setOfAssociatedRequestsWithIndices{q}{\eta'} \QSPreferenceWithIndex{q'}\setOfAssociatedRequestsWithIndices{q'}{\eta}$.
\end{definition}\vspace{-0.05in}
%By applying this definition of swap stable matching, the result will be a local maximum of the sum of the individual utilities \cite{BodineBaron2011swapMatching}.

%Solving the matching game from Definition~\ref{def:matching_game} is challenging as the QSs' preference relation consider sets of requests. Furthermore, instead of having a fixed number of requests per QS, which is called a quota in classical matching games, the limited link-level EPR pairs have to be considered for each Tx and Rx end node. 
%

\begin{figure}[tb]
\vspace*{-2.49mm}
	%\removelatexerror
	\begin{algorithm}[H]
        
		\caption{Request-QS Association (RQSA)\label{alg:matching}}
		\begin{algorithmic}[1]
			\begin{scriptsize}
			\REQUIRE Set of requests $\mathcal{R}$, set of QSs $\mathcal{Q}$.
            
            \textbf{Phase 1: Initialization Phase}
            \STATE Each request $\request$ determines its worst-case fidelities from~\eqref{eq:requestPreferences}
            \STATE Match each request $\request$ to the QS $q$ with the highest worst-case fidelity as long as constraints~\eqref{eq:constraint1} and~\eqref{eq:constraint2} are satisfied.
            %\REPEAT
            %    \STATE Each unmatched request proposes to the QS $q^*$ on top of its preference list and removes $q^*$ from its preference list
                %\STATE QS $q^*$ solves $\mathcal{P}1$ from~\eqref{eq:optProb}
            %    \STATE \textbf{if} $U_{q^*}(\mathcal{R}_{q^*} \cup \{ \request \}) > U_{q^*}(\mathcal{R}_{q^*}) $ \textbf{then} $\eta \leftarrow \eta \cup (\request, q^*)$
            %\UNTIL{all requests are matched to a QS or their preference lists are empty}
            
            \textbf{Phase 2: Swap Matching Phase}
            \REPEAT
    			\FORALL{ $r \in \mathcal{R}$ }
    			    	\STATE Select a QS $q'$ that yields a higher utility then the currently matched QS $q$
                        \FORALL{Requests $r'$ matched to QS $q'$, i.e., $r'\in\setOfAssociatedRequestsQPrime$}
                        \STATE QS $q'$ identifies a request $r'$ that shares the same Tx or Rx node with request $r$ which is not matched to $q'$
                            \IF {$q \succ_{r'}^{\text{Req}} q'$ \textbf{and} $q' \userPreference q$}
                                \STATE Construct the swap matching $\eta' \leftarrow \eta_{r,r'}^{q}$
                                \IF{$\setOfAssociatedRequestsPrime \QSPreference \setOfAssociatedRequests$\textbf{and} $\setOfAssociatedRequestsQPrimeEtaPrime\QSPreferenceWithIndex{q'} \setOfAssociatedRequestsQPrime$ \textbf{and} \eqref{eq:constraint1},\eqref{eq:constraint2} are satisfied }
                                    \STATE The swap of $r'$ and $r$ is approved, i.e., $\eta \leftarrow \eta'$
                                \ELSE
                                    \STATE The swap of $r'$ and $r$ is denied.
                                \ENDIF
                        \ENDIF
                    \ENDFOR
    			\ENDFOR
			\UNTIL{no more pairs of requests $r,r'$ to swap are found}
            \STATE All QSs $q \in \mathcal{Q}$ solve $\mathcal{P}1$ from~\eqref{eq:optProb} to determine the action for each request
            
             \textbf{Stage 3: e2e EPR Pair Generation Phase} \vspace{-0.04in}
             \STATE Each QS performs its respective actions to create e2e EPR pairs according to $\eta$
			\end{scriptsize}
		\end{algorithmic}
	\end{algorithm}\vspace{-0.3in}
\end{figure}
To solve the proposed matching game, i.e. finding a swap stable matching $\eta$, a key challenge is that the preferences~\eqref{eq:QSpreferences} of the QSs do not rank individual requests, but rather sets of requests.
Accordingly, a QS $q\in\mathcal{Q}$ cannot decide whether to accept or defer an individual request. Instead, the QS has to consider all its other associated requests in $\setOfAssociatedRequests$.
Furthermore, instead of having a fixed quota at each QS, we must consider the limited quantum memory of each QS given by constraints~\eqref{eq:constraint1} and~\eqref{eq:constraint2}.
Therefore, the well-known deferred acceptance algorithm~\cite{roth1990matchingTheory} cannot be applied to this game.

To overcome these challenges, we propose a novel request-QS association (RQSA) swap matching algorithm, which is shown in Algorithm~\ref{alg:matching}.
The matching is initialized by a greedy strategy, i.e., all submitted requests are matched to QSs with the highest request utility \eqref{eq_request_tility} as long as constraints~\eqref{eq:constraint1} and~\eqref{eq:constraint2} are satisfied for each QS (lines 1 and 2).
After initialization, the swap matching phase begins.
The preference list of each request $r$ is calculated and a more preferred QS $q'$ than its currently matched QS $q$ is identified (line 5). 
QS $q'$ identifies a request $r'$ from its associated requests that shares the same Tx or Rx node with another request $r$ not associated to $q'$ (line 7).
Then swap matching $\eta_{r,r'}^{q}$ is considered, wherein $r$ will be served by $q'$ instead of $q\in\mathcal{Q}$ and $r'$ will be served by $q\in\mathcal{Q}$ instead of $q'$ (line 8). 
The swap is performed when the requests $r$ and $r'$ and QSs $q$ and $q'$ prefer the swap, with at least one participant strictly preferring the swap matching over its current matching (line 7-16).
This procedure is repeated until no more swaps can be found in the network (line 17).
In the last stage, the e2e EPR pair generation phase, the QSs solve optimization problem $\mathcal{P}1$, identify and perform the actions to serve their associated requests (lines 18 and 19). The stability of the resulting matching $\eta$ follows from:\vspace{-0.07in}
\begin{lemma}
    Upon convergence, RQSA reaches a swap stable matching according to Definition~\ref{def:swap_stability}.
\end{lemma}\vspace{-0.05in}
\textit{Proof:} To prove swap stability upon convergence, we have to show that no pair of submitted requests $r$ and $r'$ exists with their associated QSs $q$ and $q'$, such that a swap of $r$ and $r'$ is preferred by the submitted requests and QSs. RQSA checks for all combinations of $r, r',q$ and $q'$, whether a swap is preferred by all requests and QSs. If such a combination of $r, r',q$ and $q'$ is found, the swap is performed. This procedure is repeated until no more swaps are performed. Therefore, after the swap matching phase, the resulting matching $\eta$ is swap stable as no more pairs of submitted requests remain that would prefer to be served by another QS.\footnote{Regarding the proof of convergence, we refer the reader to a general proof for swap matching algorithms in~\cite{BodineBaron2011swapMatching} due to space limitations.}

\begin{figure*}[t!]
       \centering
       \begin{minipage}[c]{0.31\linewidth}
           %\begin{center}
            \includegraphics[width=0.9\linewidth]{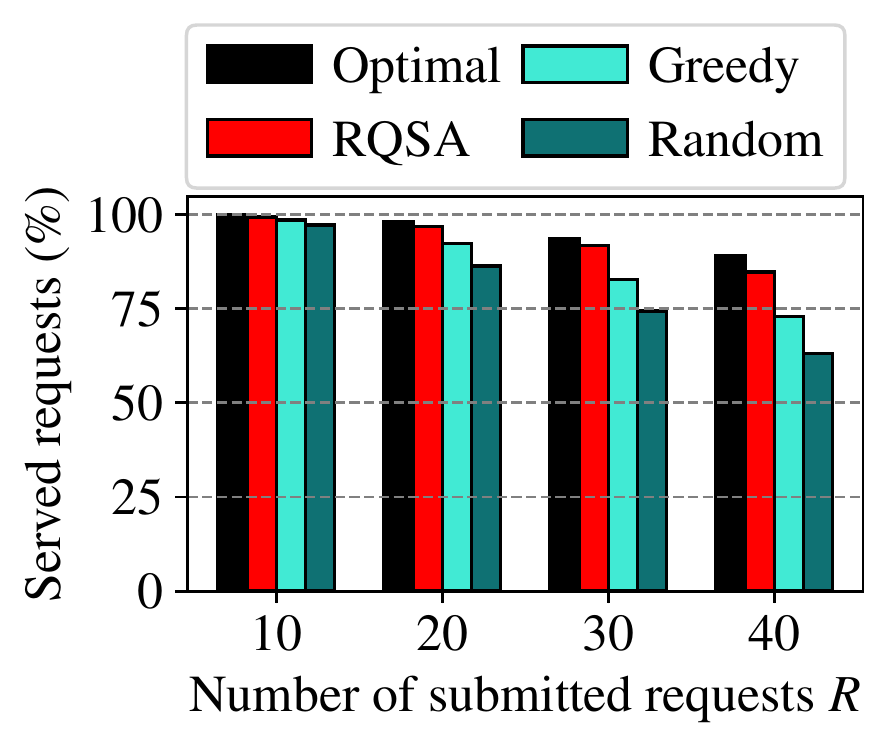}\vspace{-0.2in}
            \caption{Average percentage of served requests as a function of $R$.}
            \label{fig:subfig1}
            %\end{center}
        \end{minipage}
        \hfill
           \begin{minipage}[c]{0.31\linewidth}
             \includegraphics[width=0.9\linewidth]{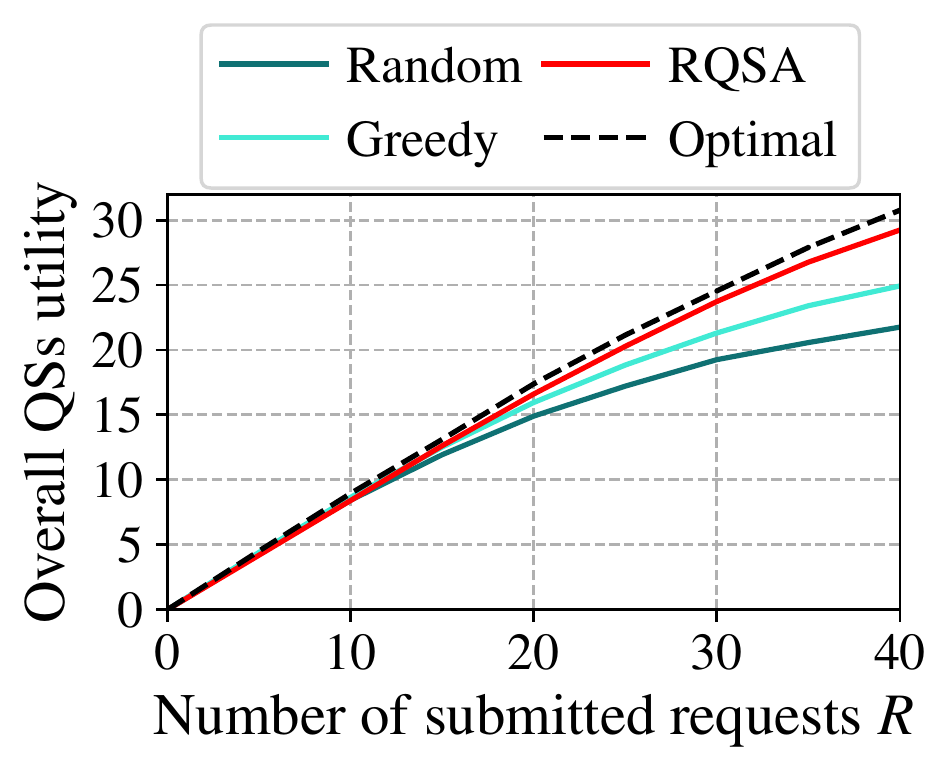}\vspace{-0.15in}
           \caption{Overall QSs' utility as a function of $R$.}
           \label{fig:subfig3} 
        \end{minipage}
        \hfill
           \begin{minipage}[c]{0.31\linewidth}
           \includegraphics[width=0.9\linewidth]{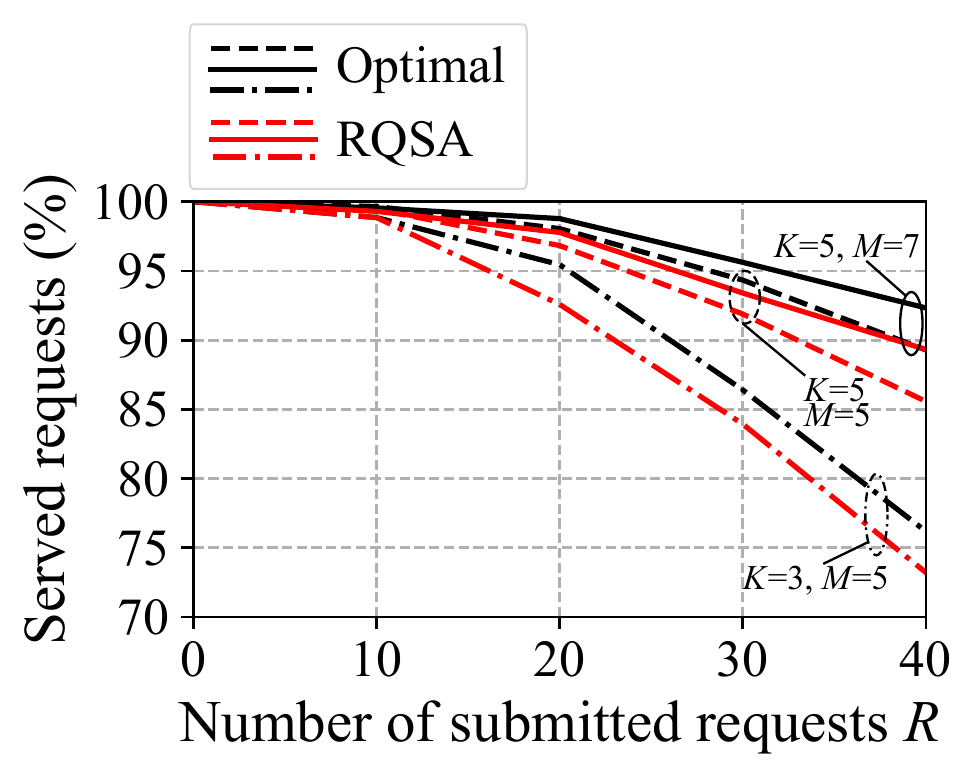}\vspace{-0.15in}
           \caption{Average percentage of served requests as a function of $K$,$M$ and $R$.}
           \label{fig:subfig4}
           %\includegraphics[width=0.9\linewidth]{figures/figure_4.pdf}\vspace{-0.15in}
           %\caption{Number of served requests as a function of $R$ and $Q$.}
           %\label{fig:subfig5}
        \end{minipage}
        \vspace{-0.25in}
    \end{figure*}

\vspace{-0.1in}    
\section{Simulation Results and Analysis}
\label{sec_experiments}\vspace{-0.06in} 
For our simulations, we define the following \emph{default} setup QCN parameters: 1) The number of Tx nodes is $K=5$, the number of Rx nodes is $M=5$, while the number of QSs is $Q=3$; 2) Heralding stations perform $n=10$ link-level EPR pair generation attempts, and the numbers of successfully-generated pairs in every time slot are binomial random variables $N_{k,q}^{\mathrm{Tx}} \sim B(n=10,p=p_{k,q})$ and $N_{q,m}^{\mathrm{Rx}} \sim B(n=10,p=p_{q,m})$ (see Sec. \ref{sec_system_model}). The probability of success is $p_{k,q} = e^{-d_{k,q}/L_0}$, for links between a QS $q\in\mathcal{Q}$ and a Tx node $k\in\mathcal{K}$, where $L_0 = \SI{0.54}{\kilo\metre}$ is the optical fiber's attenuation coefficient \cite{hensen2015loophole}, and $d_{k,q}$ is the length of those links. Similarly, $p_{q,m} = e^{-d_{q,m}/L_0}$ for links between a QS $q$ and an Rx node $m\in\mathcal{M}$. The lengths $d_{k,q}$ and $d_{q,m}$ are sampled from a uniform distribution between $\SI{100}{\metre}$ and $\SI{1}{\kilo\metre}$, $\mathcal{U}(0.1,1)$; 3) Each request has a different minimum required fidelity $F_{k,m}^{\text{min}}$ based on its intended quantum application.\footnote{Particularly, distillation protocols require a minimum fidelity of $0.5$, while $0.8$ is a typical value for quantum key distribution applications \cite{panigrahy2022capacity}.} Thus, we randomly sample such values from a uniform distribution $\mathcal{U}(0.5, 0.8)$; 4) Initial fidelities $F_{k,q}^{\mathrm{Tx}}$ and $F_{q,m}^{\mathrm{Rx}}$ of link-level EPR pairs depend on the hardware, so they are sampled from a uniform distribution $\mathcal{U}(0.83, 0.99)$ \cite{hensen2015loophole}; 5) The number of submitted requests lies in the range $R \in [0,40]$. We perform $100$ independent simulation runs wherein all aforementioned random variables are drawn from their respective distributions. Each run analyzes a single time slot where Tx nodes submit a set of requests, that we solve the request-QS association problem for. Unless stated otherwise, these default parameters are used in all simulation experiments.

% To evaluate the performance of the proposed RQSA algorithm, we consider the following baseline algorithms: 1) \emph{Optimal solution}, where the assignment of submitted requests to QSs is formulated as an integer optimization problem and solved using advanced solvers. This approach requires full QCN information to all QCN elements, which is not a practical assumption in QCNs due to its associated classical communication delay, 2) \emph{Greedy algorithm}, in which submitted requests attempt to be served by the QS with the highest worst-case fidelity. If this QS does not have enough link-level EPR pairs in its memory, the request is served by the second best QS, 3) \emph{Random algorithm}, in which each submitted request is served by a random QS. The results of the conducted experiments are discussed next. %software Baron~\cite{baronsolver}

We benchmark the proposed RQSA algorithm against the following baselines: 1) \emph{Optimal}, which formulates the request-QS association problem as an integer optimization problem solved using an advanced solver~\cite{baronsolver}. This requires complete QCN information that is impractical due to classical communication delay, 2) \emph{Greedy algorithm}, which selects the QS with the highest worst-case fidelity to serve a request, and when it lacks enough link-level EPR pairs, the next-best QS is chosen, and 3) \emph{Random algorithm}, which randomly associates each request with a QS. We discuss the experimental results next.

\subsubsection{Impact of Number of Requests on Served Requests}
First, we analyze, the performance as the number $R$ of submitted requests varies, and compare the percentage of served requests by the different algorithms. Corresponding results are shown in Fig.~\ref{fig:subfig1}, where we observe that RQSA is within $5\%$ of the optimal solution in terms of served requests. Moreover, RQSA achieves superior performance compared to both greedy and random algorithms, and the performance gap between the algorithms increases as the number of requests increases. For instance, when the number of submitted requests is $40$, RQSA serves around $13\%$ and $22\%$ more requests than the greedy and random algorithms, respectively. From Fig.~\ref{fig:subfig1}, we also observe that the optimal percentage of served requests decreases as the number of requests increases. This is due to limited QCN resources, namely link-level Tx/Rx EPR pairs, which leads to many submitted requests becoming infeasible as more requests are submitted.

\subsubsection{Impact of Number of Requests on Overall QS Utility}
Next, we show the effect of $R$ on the overall achieved QSs' utility, i.e., \emph{sum of served e2e fidelities}, in Fig.~\ref{fig:subfig3}. We observe from Fig.~\ref{fig:subfig3} that RQSA achieves near-optimal performance, even for large $R$, e.g., $R=40$. In such cases, RQSA achieves a performance within $5\%$ of the optimal overall utility, unlike the greedy and random algorithms that start to diverge from the optimal solution as $R$ becomes large. Note that, in contrast to the optimal solution algorithm, RQSA requires significantly smaller run time, and does not require full QCN information availability while being scalable.

\subsubsection{Impact of QCN Size on Performance}
Finally, in Fig.~\ref{fig:subfig4}, we analyze the scalability of RQSA by showing the percentage of served requests as $R$ varies while considering three different QCN sizes. In particular, we consider the cases in which $K>M$, $K=M$, and $K<M$ for a fixed number of QSs $Q=3$. From Fig.~\ref{fig:subfig4}, for small QCNs, e.g., $K=3$, we observe that a small number of Tx nodes imposes a bottleneck on the maximum number of served requests, since the number of available link-level EPR pairs becomes insufficient to satisfy the increased number of requests. Additionally, we observe from Fig.~\ref{fig:subfig4} that RQSA is scalable across different (small and large) QCN sizes, and it achieves a near-optimal performance, that is within $4\%$ of the optimal solution.

\vspace{-0.05in}
\section{Conclusion}\vspace{-0.05in}
\label{sec_conclusion}
In this paper, we have studied the problem of requests-QSs association in QCNs with multiple QSs, which is crucial for QCN applications like quantum data centers. To develop a practical solution and overcome the challenges of partial information and the combinatorial complexity of the association problem, we have formulated the problem as a matching game. The proposed formulation takes into account practical QCN considerations such as limited memory capacity, heterogeneous fidelity requirements, and scheduling of QS operations. Moreover, we have developed a novel swap-matching based RQSA algorithm to solve the matching game while achieving stability. Simulation results show that the proposed approach is scalable and achieves a near-optimal performance.
%Extensive simulations have been conducted, and results validated the near-optimal performance of the proposed RQSA algorithm compared to benchmark algorithms. Finally, the RQSA algorithms was shown to achieve a scalable performance with the QCN size. 

\vspace{-0.06in}
\begin{spacing}{0.8}
%\footnotesize
\bibliographystyle{IEEEtran}
\bibliography{References}
\end{spacing}

% that's all folks
\end{document}